\documentclass[aps,prd,preprint,groupedaddress]{revtex4-1}

\usepackage{bm}
\usepackage{slashed}
\usepackage{url}
\usepackage{float}

\usepackage{amsthm}
\usepackage{amssymb}
\usepackage{graphicx}
\usepackage{comment}
\usepackage{subfigure}

\newcommand{\be}{\begin{equation}}
\newcommand{\ee}{\end{equation}}
\newcommand{\bea}{\begin{eqnarray}}
\newcommand{\eea}{\end{eqnarray}}
\newcommand{\ps}{p_m\sigma^m}
\newcommand{\qs}{q_m\sigma^m}

\begin{document}

\title{Numerical study of the Dyson-Schwinger equations for the Wess-Zumino model}

\author{Marco Frasca}\email{marcofrasca@mclink.it}
\affiliation{via Erasmo Gattamelata, 3, 00176 Roma, Italy}

\begin{abstract}
Supersymmetric models, in most cases, suffer from the lack of non-perturbative techniques. Recently,
an approach based on Dyson-Schwinger equations has been proposed
for the massless Wess-Zumino model. In this case, the equations for the
self-energies of the fields were solved with the strong ansatz to take them all equal. We show, by 
numerically solving the equations, that
this is a too strong choice as also solutions with all different self-energies are acceptable and more
generic. This could have interesting implications for supersymmetry breaking. 
\end{abstract}

\date{\hfill}

\maketitle

\section{Introduction}

Studies of supersymmetric models at large coupling are rather involved and lack from
satisfactory techniques to extract analytical results. Exception for this is the use of 
AdS/CFT correspondence initially conjectured by Maldacena \cite{Maldacena:1997re} but this cannot be
helpful for a Wess-Zumino model as this symmetry applies for gauge theories. On the other side, 
supersymmetry is broken on the lattice and so lattice computations have not yielded useful insight 
so far. The problem here is to recover a meaningful limit in the continuum (see \cite{Feo:2013faa} and Refs. therein).

It is important to have a way to manage quantum field theories non-perturbatively as one can draw
definite conclusions on all momenta ranging from infrared to ultraviolet. Wild differences in the
behavior of the theories in these regimes are expected. Typical is the case of a Yang-Mills theory
displayng bounded states in a non-perturbative regime but having asymtptotic freedom on the other case
\cite{Lucini:2004my,Chen:2005mg}.

The quest for the study of non-perturbative regimes in quantum field theory has a powerful
tool provided by Dyson-Schwinger equations. Indeed, the idea
to use Dyson-Schwinger equations for the Wess-Zumino model is not completely new \cite{Bashir:1999ui}.
Recently, a new way to manage these equations for the self-energies was given and
exact results were provided for the computation of the anomalous dimension in ordinary and
supersymmetric quantum field theories \cite{kreimer1,CK1,CK2,CK3,BrKr02,BrKr01,Bellon:2007xk,Bellon:2008zz,Bellon:2013sya}. 
Particularly, the article \cite{Bellon:2007xk} has been one of the main motivations to attack the problem
in the way provided in this paper. The other main reason is that the analysis by Witten on the Wess-Zumino
model  \cite{witten82} does not seem to apply straightforwardly to our case with conformal symmetry and in the
strong coupling limit. The reason is that, in the infrared limit, the massless theory can display a spectrum of
bound states that evades Witten's analysis on this model: The breaking mechanism can be different
due to the presence of a zero mode \cite{Frasca:2013tma}. Indeed, the analysis, in the conformal
case and in the strong coupling limit has never been seen in literature. Nobody was able to check this in real life
as non-perturbative techniques are lacking. Recently, we were able to provide, using techniques developed for Yang-Mills
theory, both classical and quantum solutions \cite{Frasca:2012aj,Frasca:2012ne}. Classical solutions seem to
indicate that, when the coupling becomes very large, supersymmetry would be broken. It remains to be
seen if quantum corrections can fix the situation.

On the other side, an approach using Dyson-Schwinger equations is readily amenable to a numerical treatment. 
So, we are in a position to provide a fully numerical solution for the self-energies of the Wess-Zumino
model as in \cite{Bellon:2007xk} the equations were fully exposed and successfully applied to the
computation of the anomalous dimension in this case. But these authors made a well-definite choice
for the solution of the Dyson-Schwinger equations starting with all self-energies being equal. This is
a too strong {\sl ansatz} and must be supported by the fact that the equations, also with
a different choice, should drive to a solution like this. Equal self-energies grant that
supersymmetry is never broken. So, the main question we answer in this paper is: Is there an unique
choice for the self-energies of a massless Wess-Zumino model?

The interesting result we obtain is that the question has a negative answer. Different choices with
different self-energies for each involved field do not drive to the solution selected in \cite{Bellon:2007xk} 
and the question is somewhat more delicate. The paper is structured as follows. In Sec.\ref{sec1} we
present the model and all the notational matter. In Sec.\ref{sec2} we yield the Dyson-Schwinger equations
and put them into a numerical amenable form. In Sec.\ref{sec3} we present the numerical results. Finally,
in Sec.\ref{sec4} we give the conclusions.

\section {The model}
\label{sec1}

In all this paper the main reference is \cite{Bellon:2007xk}. So, in this section and the following one
we take the presentation of the model and the equations we started from from that article. We do this
for the sake of readers' convenience.

Our fields are given by the massless chiral superfields $\Psi$ and $\Phi_i$
($i=1,2,\ldots, N$) and their (antichiral) complex conjugates
$\Psi^+$ and $\Phi_i^+$,
with the constarints:%
\begin{eqnarray}
\bar D_{\dot \alpha} \Psi= 0 \, , \;\;&&\;\; \bar D_{\dot \alpha} \Phi_i= 0 \nonumber\\
 D_{ \alpha}   \Psi^+= 0 \, , \;\;&&\;\;   D_{ \alpha} \Phi_i^+= 0
\end{eqnarray}
being
\begin{equation}
D_\alpha = \frac{\partial}{\partial \theta^\alpha} +
 2i\sigma_{\alpha \dot\alpha}^\mu{\bar \theta}^{\dot\alpha}
 \frac{\partial}{\partial y^\mu} \; , \;\;\; \;\;\; \;\;\;
 \bar D_{\dot \alpha} = -\frac{\partial}{\partial {\bar\theta}^{\dot\alpha}}.
\end{equation}
%
%

Each chiral superfield represents a complex scalar ($A, B_i$),
a Weyl fermion ($\chi, \xi_i$) and a complex auxiliary field ($F,G_i$)
as it can be seen from their expansion in the $\theta$ variables,
\begin{eqnarray}
\Psi &=& A(y) + \sqrt 2 \theta \chi(y) + \theta \theta F(y) \nonumber
\\
\Phi_i &=& B_i(y) + \sqrt 2 \theta \xi_i (y) + \theta \theta G_i(y).
\label{uno}
\end{eqnarray}

The Lagrangian density is
\begin{equation}
L = \int d^4\theta \Psi \Psi^+ +
\sum_{i=1}^N \int d^4\theta \Phi_i \Phi_i^+
+
\frac{g}{\sqrt N} \sum_{i=1}^N \int d^2\theta \Phi_i\Psi^2.
\end{equation}
%
that written in terms of component fields is
\begin{eqnarray}
L &=& i\partial_\mu \bar \chi \bar \sigma^\mu \chi +
i\sum_{i=1}^N \partial_\mu \bar \xi_i \bar \sigma^\mu \xi_i
+ A^* \Box A + \sum_{i=1}^N  B_i^* \Box B_i +
F^*F + \sum_{i=1}^N G_i^*G_i
\nonumber\\
&& + \frac{g}{\sqrt N} \sum_{i=1}^N \left(
 A^2 G_i +2 A B_i F- \chi^2 B_i - 2 \xi_i\chi A +{\rm h.c.}
\right)
\end{eqnarray}
Finally, we take ${\rm diag}\; \eta_{\mu\nu} = (-1,1,1,1)$ and
\begin{equation}
\Box = \eta_{\mu\nu}\partial^\mu\partial^\nu=\partial_\mu\partial^\mu= -\partial_t^2 + \nabla^2.
\end{equation}

\section{Dyson--Schwinger equations}
\label{sec2}

The Dyson-Schwinger equations for the propagators of the superfields
we will work with are derived in \cite{Bellon:2007xk}. One has
\begin{eqnarray}
\Pi_{0A}^{-1}(p) &=& \langle 0 \vert T\left( A(x)
A^*(x')\right)\vert 0\rangle = i\,
\Box^{-1}(x-x')\nonumber\\
\Pi_{0F}^{-1}(p)&=& \langle 0 \vert T\left( F(x) F^*(x')\right)\vert
0\rangle =
  i\,
\delta(x-x')\nonumber\\
\left(\Pi_{0\chi }^{-1}\right)_{\!\!\alpha \dot \beta}\!\!(p) &=& \langle 0 \vert T\left( \chi_\alpha(x)
{\bar\chi}_{\dot\beta}(x')\right)\vert 0\rangle = \sigma^m_{\alpha
\dot\beta}  \partial_m \Box^{-1}(x-x')
\end{eqnarray}Moving to the Euclidean space with $\Pi_F$, $\Pi_\chi$, $\Pi_A$ the full
propagators for fields $F, \chi$ and $A$ in (Euclidean) momentum
space, the corresponding self-energies are defined to be
\begin{eqnarray}
\Pi_A^{-1}(p) &=& p^2 \left(1 - \Sigma_A(p^2)\right)\nonumber\\
\Pi_F^{-1}(p) &=& -(1 - \Sigma_F(p^2)) \nonumber\\
\Pi_\chi^{-1}(p) &=& p_m\sigma^m \left(1 - \Sigma_\chi(p^2)\right)
\end{eqnarray}
The corresponding self-energies at one-loop are 
\begin{equation} \label{sigma}
    \Sigma_F(p^2) = -\frac{g^2}{4\pi^4}\int d^4q {1\over q^2 (1-\Sigma_A(q^2)) (p-q)^2}
\end{equation}
for the auxiliary field,
\begin{eqnarray}
\ps \, \Sigma_\chi(p^2) &=&  -\frac{g^2}{4\pi^4}\left(
\int d^4q \frac{\qs}{q^2(1- \Sigma_\chi(q^2))(p-q)^2} + \right.\nonumber\\
&& \left.\int d^4q \frac{\ps - \qs}{q^2(1- \Sigma_A(q^2))(p-q)^2}
\right)\label{sigmaChi}
\end{eqnarray}
for the fermion field and
\begin{eqnarray}
p^2 \Sigma_A(p^2) &=&  \frac{g^2}{4\pi^4} \left(
\int d^4q \frac{-1}{q^2(1- \Sigma_A(q^2))}
+ \int d^4q \frac{-1}{(1- \Sigma_F(q^2))(p-q)^2}\right.
\nonumber\\
&+&\left.
\int d^4q \frac{-{\rm Tr}\bigl(q_n\bar\sigma^n(\qs-\ps) \bigr)}
{q^2(1- \Sigma_\chi(q^2))(p-q)^2} \right)\label{sigmaA}
\end{eqnarray}
for the scalar field

In order to solve these equations and consistently with supersymmetry, in
\cite{Bellon:2007xk} authors have chosen
\begin{equation}
\label{eq:ansatz}
\Sigma_A = \Sigma_\chi = \Sigma_F = \Sigma
\end{equation}
This is the key point. Indeed,
doing this, the two integrals in~(\ref{sigmaChi}) combine
to give $\ps$ times the right hand side of~(\ref{sigma}). Similarly,
the equation for $\Sigma_A$~(\ref{sigmaA}) becomes $p^2$
times~(\ref{sigma}).

The integration in (\ref{sigma}) can be done along the same lines
as in~\cite{BrKr01}. The angular integration uses the fact that
the angular average of $1/(p-q)^2$ is $1/\max(p^2,q^2)$ and we note the
fundamental identity
\begin{equation}
    \left\langle\frac{p\cdot q}{(p-q)^2}\right\rangle=\frac{1}{2}\frac{p^2+q^2}{\text{max}(p^2,q^2)}-\frac{1}{2}.
\end{equation}

With the variables $x=p^2$ and $y=q^2$, the radiative corrections become ($\lambda=g^2/4\pi^2$)
\begin{eqnarray}
\label{eq:ds}
   \Sigma_F(x)&=&\lambda\int_{\mu^2}^x\frac{dy}{y(1-\Sigma_A(y))}+\tilde F(\mu^2)-\tilde F(x) \nonumber \\
   \Sigma_\chi(x)&=&\Sigma_F(x)+\frac{\lambda}{2}
   \int_{\mu^2}^x\frac{dy}{y}\left[\frac{1}{1-\Sigma_\chi(y)}-\frac{1}{1-\Sigma_A(y)}\right]+ \nonumber \\
   &&\Xi(\mu^2)-\Xi(x) \nonumber \\
   \Sigma_A(x)&=&\lambda\int_{\mu^2}^x\frac{dy}{y(1-\Sigma_\chi(y))}+{\tilde F}(\mu^2)-{\tilde F}(x)+A_1(\mu^2)-A_1(x)+A_2(\mu^2)-A_2(x)
\end{eqnarray}
where we have used the renormalization condition $\Sigma_F(\mu^2)=\Sigma_\chi(\mu^2)=\Sigma_A(\mu^2)=0$ and
\begin{eqnarray}
   \tilde F(x)&=&\lambda\int_0^x\frac{dy}{x(1-\Sigma_A(y))} \nonumber \\
   \Xi(x)&=&\frac{\lambda}{2}\int_0^xdy\frac{y}{x^2}\left[\frac{1}{1-\Sigma_\chi(y)}-\frac{1}{1-\Sigma_A(y)}\right] \nonumber \\
   A_1(x)&=&-\lambda\int_0^x dy\frac{y}{x^2}\left(\frac{1}{1-\Sigma_\chi(y)}-\frac{1}{1-\Sigma_F(y)}\right) \nonumber \\
   A_2(x)&=&-\lambda\int_x^\infty\frac{dy}{x}\left[\frac{2}{1-\Sigma_\chi(y)}-\frac{1}{1-\Sigma_F(y)}-\frac{1}{1-\Sigma_A(y)}\right].
\end{eqnarray}
This is consistent with the solution proposed in \cite{Bellon:2007xk} eq.(\ref{eq:ansatz}) $\Sigma_F=\Sigma_A=\Sigma_\chi=\Sigma$. Our aim is to see, by numerically solving the above equations, if the ansatz chosen in \cite{Bellon:2007xk} is unique. We just note that such a choice for a solution prevents any possible breaking of supersymmetry. In any case, self-energies never get a constant value.

\section{Numerical results}
\label{sec3}

In order to check for the uniqueness of the choice of the self-energies in \cite{Bellon:2007xk} we implemented numerically eq.(\ref{eq:ds}) with an iterative procedure. This technique is well-known and very simple to adopt. In order to verify the consistency of our approach we need to check that the final result does not depend on the choice of the first iterate and that all the procedure converges. This last control was operated by varying the number of iterations.

\begin{figure}[H]
\begin{center}
\subfigure[]{
\includegraphics[angle=0, width=.45\textwidth]{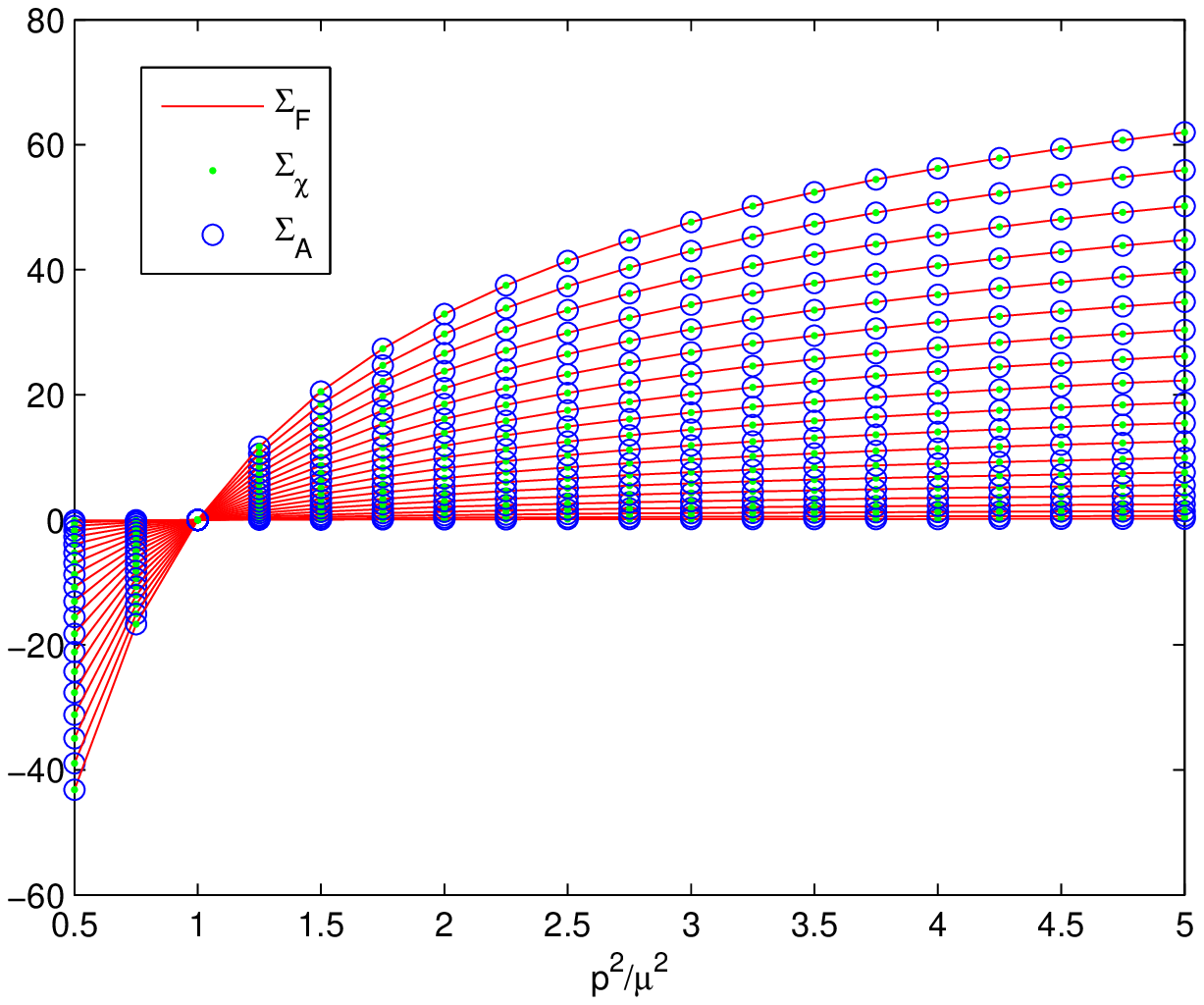}
}
\subfigure[]{
\includegraphics[angle=0, width=.45\textwidth]{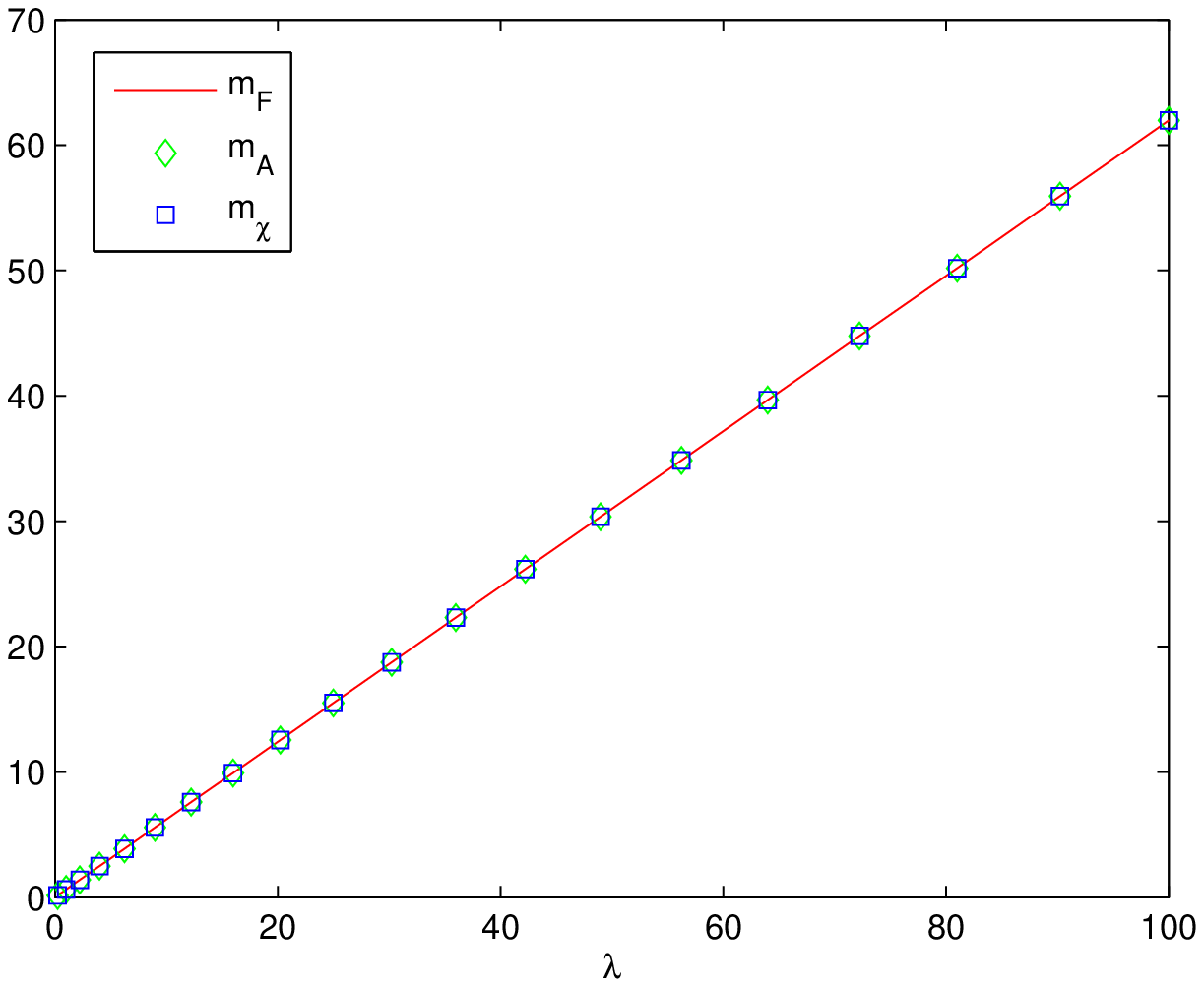}
}
\caption{\label{fig:eq} Simulation with all the self-energies being equal. (a) Self-energies as a function of momenta at different couplings (from 0.25 to 100). (b) Last values of the self-energies at different couplings. It is evident the direct proportionality.}
\end{center}
\end{figure}

The first step was to check the {\sl ansatz} of the authors in \cite{Bellon:2007xk} starting the iteration in the numerical code with all the self-energies assumed to have an equal functional form. The iterative procedure converges after a few steps to the proper values, as can be seen in Fig.\ref{fig:eq}. Independently on the value of the coupling, the self-energies keep on stay on identical values. We have also numerically checked that $(1-\Sigma)D(D+1)\Sigma$, being $D=xd/dx$, settles to a constant and this is indeed the case.

We observed that this result is fully independent from the choice of the first iterate in the procedure and the Dyson-Schwinger equations indeed settle on a solution with all equal self-energies when this ansatz is set from the start.

\begin{figure}[H]
\begin{center}
\subfigure[]{
\includegraphics[angle=0, width=.45\textwidth]{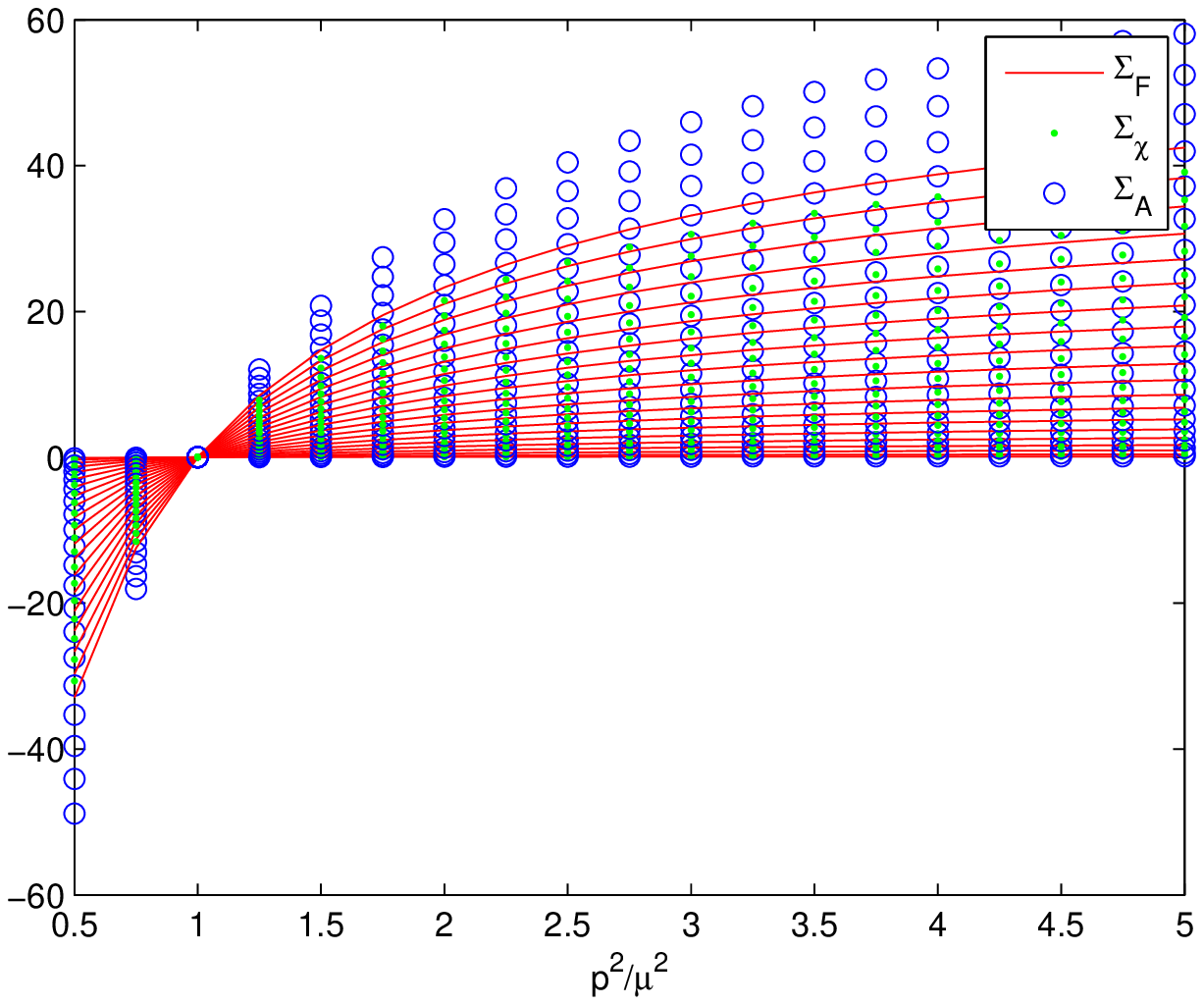}
}
\subfigure[]{
\includegraphics[angle=0, width=.45\textwidth]{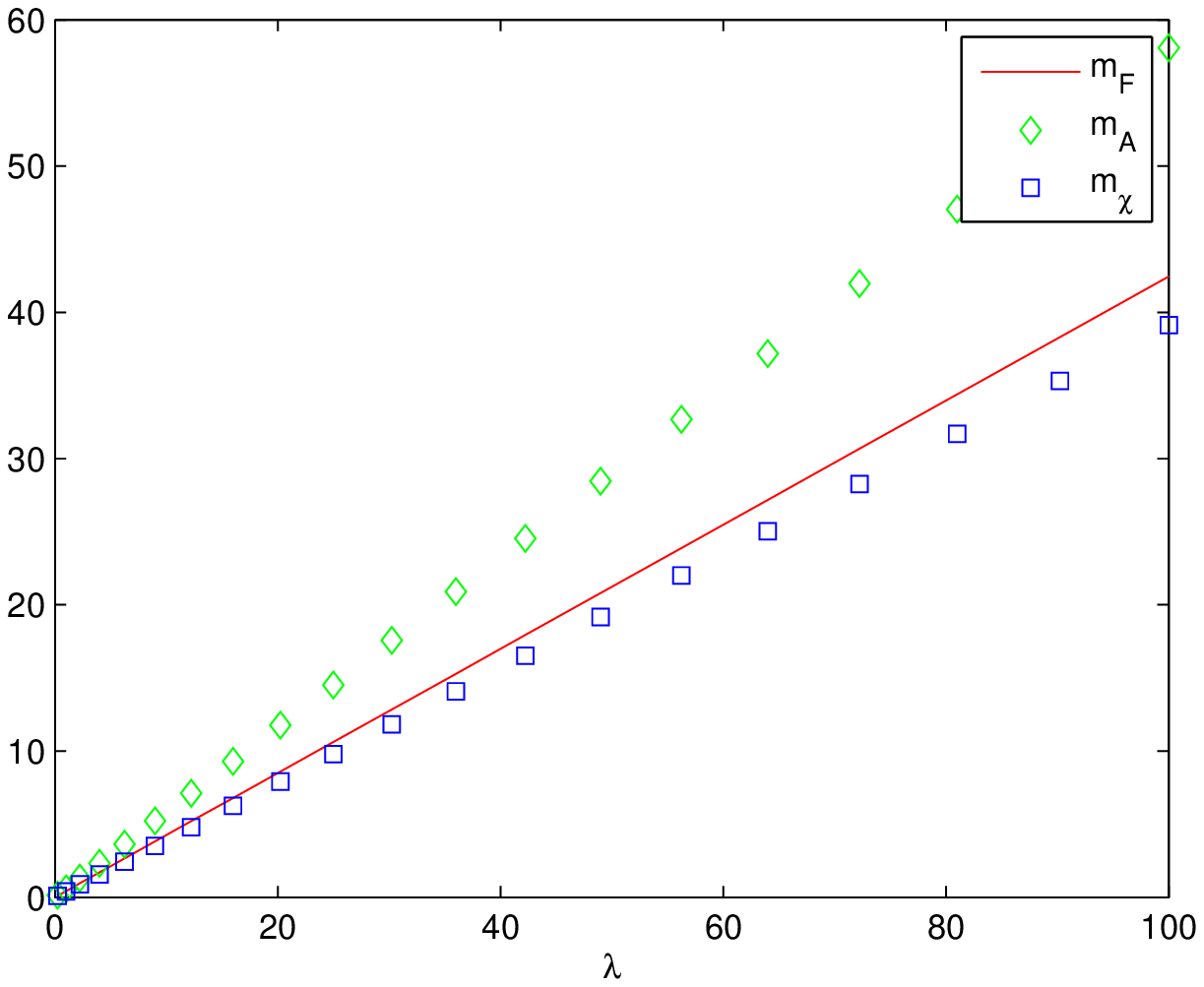}
}
\caption{\label{fig:neq} Simulation with different self-energies as first iterate. (a) Self-energies as a function of momenta at different couplings (from 0.25 to 100). (b) Last values of the self-energies at different couplings. Again, direct proportionality is seen.}
\end{center}
\end{figure}

In order to answer the question of uniqueness we have to see what happens to Dyson-Schwinger equation when different values of the self-energies are chosen as first iterate deviating in this way from the original {\sl ansatz} of the authors in \cite{Bellon:2007xk}. We performed this task and observed that in this case the solutions do not settle at all at the same value unless small couplings are considered. So, the particular choice of equal self-energies is not a generic one for the Wess-Zumino model. This can be seen at glance in Fig.\ref{fig:neq}

The effect of a different choice of a first iterate introduces a constant scale factor into the computation of the self-energies independently on the way these are chosen (if equal or not) without any deviation from the functional form implied. It is also interesting to note the behavior of the self-energies at different couplings: For small couplings these are almost well preserving supersymmetry that appears broken at strong couplings instead, a behavior that appears to be almost generic.

\section{Conclusions}
\label{sec4}

The numerical solution of Dyson-Schwinger equations for the self-energies of the Wess-Zumino model yields some unexpected results. These equations admit a simple solution assuming the self-energies are equal for all the fields in the model. But we have shown that this choice is not a generic one as this model gets solutions also with different choices. This could have far-reaching consequences unless a more mundane explanation is found. We hope to extend this analysis further for a better understanding of these results.

\end{document}